\documentclass{article}

\usepackage[margin=1in]{geometry}
\usepackage[utf8]{inputenc} 
\usepackage[T1]{fontenc}    
\usepackage{hyperref}       
\usepackage{url}            
\usepackage{booktabs}       
\usepackage{amsfonts}       
\usepackage{nicefrac}       
\usepackage{microtype}      
\usepackage{lipsum}
\usepackage{graphicx}
\usepackage{setspace}
\usepackage{longtable}
\usepackage{color}

\usepackage[numbers,comma, square]{natbib}
\usepackage{amsmath,amssymb,amsthm} 

\usepackage{abstract}
\usepackage{authblk}

\title{High efficacy of layered controls for reducing exposure to airborne pathogens}

\author[1,*]{Laura Fierce\thanks{To whom correspondence should be addressed. Email: laura.fierce\@pnnl.gov}}
\author[2]{Alison J. Robey}
\author[3]{ Cathrine Hamilton }
\affil[1]{Atmospheric Sciences and Global Change Division, Pacific Northwest National Laboratory, Richland, Washington, USA}
\affil[*]{Previously at: Environmental \& Climate Sciences Department, Brookhaven National Laboratory, Upton, New York, USA}
\affil[2]{Center for Environmental Studies, Williams College, Williamstown, Massachusetts, USA}
\affil[3]{Department of Chemistry, Indiana University of Pennsylvania, Indiana, Pennsylvania, USA}

\begin{document}

\maketitle
\begin{abstract}
To optimize strategies for curbing the transmission of airborne pathogens, the efficacy of three key controls --- face masks, ventilation, and physical distancing --- must be well understood. In this study we used the Quadrature-based model of Respiratory Aerosol and Droplets to quantify the reduction in exposure to airborne pathogens from various combinations of controls. For each combination of controls, we simulated thousands of scenarios that represent the tremendous variability in factors governing airborne transmission and the efficacy of mitigation strategies. While the efficacy of any individual control was highly variable among scenarios, combining universal mask-wearing with distancing of 1~m or more reduced the median exposure by more than 99\% relative to a close, unmasked conversation, with further reductions if ventilation is also enhanced. The large reductions in exposure to airborne pathogens translated to large reductions in the risk of initial infection in a new host. These findings suggest that layering controls is highly effective for reducing transmission of airborne pathogens and will be critical for curbing outbreaks of novel viruses in the future.
\end{abstract}

\section*{Practical Implications}
\begin{itemize}
    \item The combination of physical distancing, increased ventilation, and use of face masks by both parties reduced the risk of infection by more than 98\% in 95\% of model cases, suggesting that these nonpharmaceutical interventions offer important layers of protection when the risk of airborne transmission is high.
    \item Increasing the ventilation rate by 4~ACH led to a median reduction in the risk of infection by 90\% if the infections and susceptible person were distanced by 2~m, whereas the impact of ventilation on transmission rates was negligible during close, face-to-face interactions.
    \item Variability in the efficacy of face masks in reducing transmission is driven almost entirely by the mask collection efficiency, indicating that public health policy should emphasize the use of high-efficiency masks, such as N95 respirators. 
\end{itemize}

\section{Introduction}\label{sec:intro}
In the past two decades, novel viruses capable of airborne transmission have emerged with alarming frequency, including SARS-CoV-1 in 2003 \citep{yu2004}, H1N1 in 2009 \citep{jayaraman2011}, MERS in 2012 \citep{kim2016}, and SAR-CoV-2 in 2019 \citep{greenhalgh2021}. Yet, in many countries, controls on airborne transmission were not widely adopted until the COVID-19 pandemic \citep{bahl2020}. SARS-CoV-2, the virus that causes COVID-19, is spread when virus-laden particles that were expelled from an infectious are inhaled by a new host \citep{greenhalgh2021, samet2021, prather2020}. The risk of infection in the new host depends on the number of virions reaching the infection site and their immune responses \citep{gale2020}. Though SARS-CoV-2 is often contained in particles small enough to remain suspended and contribute to long-range airborne transmission \citep{li2021, miller2021, azimi2021}, the concentration of airborne pathogens is orders of magnitude greater directly downwind of an infectious individual than in the larger indoor space \citep{chen2020}. In this study, we quantify how different combinations of nonpharmaceutical interventions, including face masks, ventilation, and distancing, affect exposure to airborne pathogens through short-range and long-range airborne transmission.

Factors driving airborne transmission of pathogens and the efficacy of a particular mitigation strategy are inherently variable and, often, poorly constrained \citep[][and references therein]{fierce2021IndoorAir}. For example, while the size-resolved collection efficiencies of different types of face masks have been measured \citep{pan2021}, the impact of face masks on the risk of infection may also be affected by factors such as rates of viral shedding \citep{leung2020}, characteristics of expiratory jets \citep{tang2013}, room conditions \citep{qian2018}, and immune responses \citep{watanabe2010}, all of which are highly variable \citep[][and references therein]{fierce2021IndoorAir}. This uncertainty can be represented in models by simulating ensembles of scenarios, but models designed to simulate the evolution of respiratory particles in indoor spaces are often computationally expensive, limiting the number of simulations that can be performed. 

Here we use a mechanistic model to quantify the efficacy of individual and combined controls in reducing exposure to airborne pathogens. We applied the new Quadrature-based Model of Respiratory Aerosol and Droplets (QuaRAD) \citep[][]{fierce2021IndoorAir} to simulate 4,000 scenarios, which represent uncertainty in factors governing airborne transmission. QuaRAD uses a quadrature representation of the aerosol size distribution to simulate the life cycle of pathogen-laden particles, from their initial creation in the respiratory system to their eventual removal from the room or deposition into the respiratory system of a new host. The quadrature-based representation offers a balance between accuracy and computational efficiency that enables simulation the combination of short-range and long-range airborne transmission across many different scenarios. Here we used QuaRAD quantify the impact of various combinations of nonpharmaceutical interventions on the risk of initial infection in a susceptible person during a face-to-face conversation with an infectious person, while also accounting for the inherent variability in factors governing transmission risk. An overview of the modeling approach is described in Section~\ref{sec:methods}; results are presented in Section~\ref{sec:results}; and a discussion of the results and conclusions are presented in Section~\ref{sec:conclusions}.

\section{Methods}\label{sec:methods}
\subsection{Quadrature-based model of Respiratory Aerosol and Droplets\label{sec:model_description}}
In this study, we used QuaRAD to simulate the risk of initial infection in a susceptible person during a face-to-face conversation with an infectious person --- and the reduction in that risk with various combinations of nonpharmaceutical interventions. We focused on asymptomatic or presymptomatic transmission only, with each scenario representing an infectious person speaking continuously in an indoor space for three hours.

QuaRAD is designed to efficiently simulate the lifecycle of respiratory aerosol and droplets within indoor spaces. Aerosol and droplet dynamics in QuaRAD are simulated using the Quadrature Method of Moments \citep{mcgraw1997}, which replaces the continuous aerosol distribution with a small set of quadrature points and associated number concentrations. Originally developed for simulation of atmospheric aerosol, quadrature representations of aerosol distributions have been shown to reproduce key integrals over the continuous distribution with high accuracy \citep{mcgraw1997,fierce2017}. The model represents respiratory aerosol and droplets with three superimposed lognormal size distributions that correspond to their site of origin in the respiratory system, based on the framework introduced by \cite{morawska2009,johnson2011}. We previously showed that exposure to airborne pathogens are accurately represented using a set of six weighted particles, corresponding to 1-point, 3-point, and 2-point quadrature to represent particles from the bronchial, laryngeal, and oral regions, respectively (see Appendix 1 of \cite{fierce2021IndoorAir}). The size distribution of expired aerosol and, thereby, quadrature approximation is varied among scenarios.

After pathogen-laden particles are expelled, QuaRAD simulates their transport through the expiratory jet of an infectious person and within the larger indoor space, changes in particle size through water evaporation, pathogen inactivation within particles, and particle removal through ventilation, deposition to surfaces, and gravitational settling. Here we provide an overview of the QuaRAD model, with a complete description provided in \cite{fierce2021IndoorAir}. Distributions in model input parameters are provided in Table~1 of \cite{fierce2021IndoorAir}.

\subsubsection{Evaporation and pathogen inactivation}
We assume that particles are expelled from an infectious person in air that is saturated with water vapor (RH=100\%~$\pm~0.5$\%) and at the average human body temperature (310.15~K~$\pm~0.1$~K). The temperature and relative humidity rapidly decrease as particles are transported within the jet and throughout the broader indoor space, leading to water evaporation from the particle. The rate at which water is driven from the particle and its final equilibrium size is modeled as a function of the indoor relative humidity and temperature, as well as the particle's size and composition. The environmental properties within the indoor space and particle characteristics were varied among scenarios, with distributions in input parameters provided in Table~1 of \cite{fierce2021IndoorAir}.

We modeled the temporal evolution of each quadrature point using the evaporation model of \cite{kukkonen1989} as implemented in \cite{wei2015} by solving the following set of coupled ordinary differential equations:
\begin{align}\label{eqn:evaporation}
&\frac{dm_{\text{p}}}{dt}=\frac{2\pi pD_{\text{p}}M_{\text{w}}D_{\infty}C_{\text{T}}\text{Sh}}{RT_{\text{v},\infty}}\ln\bigg(\frac{p-p_{\text{v,p}}}{p-p_{\text{v},\infty}}\bigg)\\
&\frac{dT_{\text{p}}}{dt}=\frac{1}{m_{\text{p}}C_{\text{p}}}\bigg(\pi D_{\text{p}}^2k_g\frac{T_{\text{v},\infty} - T_{\text{p}}}{0.5D_{\text{p}}}\text{Nu}-L_{\text{v}}\frac{dm_{\text{p}}}{dt}\bigg),
\end{align}
where $m_{\text{p}}$ is the particle mass, $T_{\text{p}}$ is the particle temperature, $p$ is the ambient pressure, $p_{\text{v,p}}$ is the vapor pressure at the droplet surface, $p_{\text{v},\infty}$ is the vapor pressure of water in air near the particle, $M_{\text{w}}$ is the molecular weight of water, $D_{\infty}$ is the binary diffusion coefficient of water vapor in air far from the droplet surface, $R$ is the universal gas constant, $C_{\text{p}}$ is the specific heat of the particle, $k_{\text{g}}$ is the thermal conductivity of air, $L_{\text{v}}$ is the latent heat of vaporization, $C_{\text{T}}$ is a correction factor, Sh is the Sherwood number, and  Nu is the Nusselt number. Equations for terms in the evaporation model are provided in \cite{fierce2021IndoorAir}. Assuming aerosol particles and droplets are spherical and that the mass of non-water aerosol components remains constant over the simulation, we then compute the droplet diameter $D_{\text{p}}$.

In addition to affecting particle size, evaporation modifies the concentration of solutes within particles, which is thought to alter the rate at which virions are inactivated \citep{morris2020,marr2019}. Here we predicted pathogen inactivation using inactivation timescale for SARS-CoV-2, as reported in \cite{morris2020}; the implementation of this approach in QuaRAD is described in \cite{robey2022sensitivity}.

\subsubsection{Transport and removal}\label{sec:quarad_transport}
For each of the quadrature points, we use a Gaussian puff dispersion model \citep{drivas1996} to simulate dispersion as particles are transported within the turbulent jet of the infectious person and throughout the larger indoor space. We applied the turbulent jet model of \cite{lee2003} to estimate the gas velocity, temperature, and relative humidity fields within the expiratory jet. We model the dispersion of aerosol particles and droplets from continuous speech as a series of puffs expelled in succession, the concentration of particles at a particular location ($x,y,z$) and time $t$ is then computed as the sum over all Gaussian puffs that had been expelled between the time the conversation started and $t$.

The spatiotemporal evolution of the number concentration of pathogens from an individual puff and associated with quadrature point $i$ is modeled according to \cite{drivas1996}:
\begin{equation}
n_{\text{v},i}(x,y,z,t) = \frac{N_{\text{v},0}\exp{(-\lambda t)}}{\pi^{3/2}b^3}R_x(t)R_y(t)R_z(t),
\end{equation}
where $t$ is the time since the puff was expired, $N_{\text{v},0}=\dot{N}_{\text{v}}\Delta t$, is the number of pathogens associated with quadrature point $i$ that are expelled between $t_k$ and $t_k+\Delta{t}$, $\lambda$ is the pathogen removal rate through a combination of ventilation, gravitational settling, deposition to surfaces through diffusion, and inactivation,  $b(t)$ is the puff width,  and $R_x(t)$, $R_y(t)$, and $R_z(t)$ represent dispersion and reflection terms. The model from \cite{drivas1996} represents dispersion from a point source under quiescent conditions, including reflections from walls if in an enclosed space. We expanded upon this model to represent the evolution of a continuous source, represented as a series of Gaussian puffs, that are transported within an expiratory jet and throughout the larger room. In \cite{drivas1996}, the width of each puff $b(t)$ is parameterized as a function of the turbulent diffusion coefficient within the larger indoor space. In contrast, we assume that puff dispersion is driven by the turbulent gas jet created by the infectious person, such that the center of puff is transported according to the gas velocity field within the jet and larger indoor space. We assume $b(t)$ corresponds to the Gaussian half-width of the jet, which increases linearly with distance along the centerline of the jet. 

To represent dispersion within an enclosed spaces, we follow \cite{drivas1996} and include reflections from the walls, floor, and ceiling for particles smaller than $30~\mu$m; $R_x(t)$, $R_y(t)$, and $R_z(t)$ are given by:
\begin{align}
R_x(t)=\sum_{j=-\infty}^{\infty}\Bigg[\exp\bigg({-\frac{(x+2jL-x_{\text{c}}(t))^2}{b(t)^2}}\bigg) + \exp{\bigg(-\frac{(x+2jL+x_{\text{c}}(t))^2}{b(t)^2}\bigg)}\Bigg]\label{eqn:Rx}\\
R_y(t)=\sum_{j=-\infty}^{\infty}\Bigg[\exp{\bigg(-\frac{(y+2jW-y_{\text{c}}(t))^2}{b(t)^2}\bigg)} + \exp{\bigg(-\frac{(y+2jW+y_{\text{c}}(t))^2}{b(t)^2}\bigg)}\Bigg]\label{eqn:Ry}\\
R_z(t)=\sum_{j=-\infty}^{\infty}\Bigg[\exp{\bigg(-\frac{(z+2jH-z_{\text{c}}(t))^2}{b(t)^2}\bigg)} + \exp{\bigg(-\frac{(z+2jH+z_{\text{c}}(t))^2}{b(t)^2}\bigg)}\Bigg]\label{eqn:Rz},
\end{align}
where $L$, $W$, and $H$ are the length, width, and height of the room, and $(x_{\text{c}}(t),y_{\text{c}}(t),z_{\text{c}}(t))$ is the trajectory of a particle located in the center of the puff. For larger particles, wall reflections are neglected, and we include only the terms for $j=0$, and for small particle we truncate the infinite sum at 500 reflections. The centerline trajectory of each puff $(x_{\text{c}}(t),y_{\text{c}}(t),z_{\text{c}}(t))$ is computed by solving the equations of motion for particles within a moving gas. 

\subsubsection{Deposition to the respiratory system and risk of initial infection}
We assume that the risk of initial infection in a new host depends on the total number of pathogens reaching the infection site, $N_{\text{v}}$, and the risk of initial infect per pathogen reaching the infection site, $p_1$, as given in the model of \cite{gale2020}:
\begin{equation}\label{eqn:dose_response}
p_{\text{infect}}=1 - (1-p_1)^{N_{\text{v}}}.
\end{equation}
The overall pathogen dose, $N_{\text{v}}$, is the sum over $N_{\text{v},i}$ for quadrature points $i=1,...,N_{\text{quad}}$, where $N_{\text{v},i}$, the total number of pathogens associated with a given quadrature point that reach the infection site, is given by:
\begin{equation}
N_{\text{V},i} = \int_0^tn(x,y,z,t)\dot{V}_{\text{inhale}}DE(D_i)dt,
\end{equation}
where $\dot{V}_{\text{inhale}}$ is the breathing rate, which is varied among scenarios to represent the variability in $\dot{V}_{\text{inhale}}$ among adults during normal breathing, and $DE(D_i)$ is the size-dependent deposition efficiency. In the case of COVID-19, the initial infection site is the upper respiratory tract, and we estimate the pathogen dose to this region using the deposition model from \cite{cheng2003} (see eqn.~41 of \cite{fierce2021IndoorAir}), wherein parameters of the deposition model are varied among simulations to represent reported uncertainties (see Appendix of \cite{fierce2021IndoorAir}).

Although the risk of initial infection depends strongly on poorly constrained physiological parameters, such as $p_1$ and the viral load, the reduction in the risk of infection associated with controls on airborne transmission is insensitive to these parameters. Through sensitivity simulations, we found that the efficacy of controls varied with $p_1$ and $N_{\text{v}}$ only in extreme cases in which the risk of infection would be nearly 100\% even when controls are employed, such as in the case of a highly transmission pathogen or during long (>24~h), very close (<0.5~m) interactions. However, across the range of scenarios explored here, we found reductions in the risk of initial infection scaled directly with reductions in exposure to airborne pathogens ($R^2>99.9\%$).

\subsection{Sampling model input parameters}\label{sec:methods_scenarios}
To represent the wide range of conditions within indoor spaces and the wide variation in physiology among the general population, we performed 4,000 model simulations with QuaRAD. We used Latin Hypercube Sampling from the Python Design of Experiments library (https://pythonhosted.org/pyDOE/) to sample uncertainty in model input parameters, which are provided in Table~1 of \cite{fierce2021IndoorAir}. Where possible, distributions in model parameters were constrained using measurements.  

Air change rates in the baseline cases were sampled from a uniform distribution ranging from 0.3 to 2.7 air changes per hour  \cite{epa2018,bennett2012,turk1987}. The length and width of the room were sampled from uniform distributions ranging from 7~m to 10~m \cite{nystate2010}; the height of the room was sampled from a normal distribution ranging with mean of 2.74~m and standard deviation of 0.38~m \cite{gsa2019}; and the expiration velocity was sampled from a normal distribution with mean of 4~m/s and standard deviation of 2~m/s \citep{chao2009}. 

The size distribution of respiratory aerosol and droplets was represented by three superimposed lognormal modes that correspond to sites of origin in the respiratory system \citep{johnson2011,morawska2009}: the smallest in the b-mode (bronchial), mid-sized in the l-mode (laryngeal), the largest in the o-mode (oral). The size distribution parameters for each mode and, consequently, quadrature points and weights were varied among scenarios, according to the uncertainties reported in \cite{johnson2011,morawska2009}. The viral loads in particles of varying sizes is not well constrained for SARS-CoV-2, so we used virion expiration rates among particles of different sizes that have been measured for influenza \citep{milton2013, leung2020}, which has a viral shedding pattern similar to SARS-CoV-2 \citep{jacot2020}. We combined the distributions in the number concentration and size distribution parameters for particles in expired breath from \cite{johnson2011,morawska2009}, the number concentration of pathogens in air associated with fine and coarse particles from \cite{milton2013, leung2020}, and expiration rates from \cite{flenady2017} to predict the pathogen expiration rate for the b-, l-, and o-modes, each of which varies by orders of magnitude. Parameters of the model used to predict deposition to the nasal cavity of a susceptible person was also varied according to \citep{cheng2003}. See Tables~1 of \cite{fierce2021IndoorAir} for distributions in model input parameters.

\begin{table}[h!]
    \centering
 \begin{tabular}{c c c c c}
        $D_{\text{p}}$ ($\mu$m) & $\text{CE}_{\text{in}}$ & $\sigma_{\text{in}}$ & $\text{CE}_{\text{out}}$ & $\sigma_{\text{out}}$ \\ \hline
        0.5 & 0.3 & 0.2 & 0.4 & 1.2 \\
        0.7 & 0.3 & 0.2 & 0.6 & 1.0 \\
        1.0 & 0.3 & 0.2 & 0.7 & 0.8 \\
        2.0 & 0.5 & 0.2 & 0.8 & 0.7 \\
        5.0 & 0.9 & 0.1 & 0.9 & 0.2 \\ \hline
    \end{tabular}
    \caption{Mean and standard deviation ($\sigma$) of collection efficiency (CE) of a surgical mask measured in \citep{pan2021}. Surgical masks also reduce expiration velocity by an average of 53\% ($\sigma=0.12$) \citep{maher2020}.}\label{tab:maskFE}
\end{table}

\subsection{Quantifying efficacy of controls}\label{sec:methods_efficacy}

For each of the 4,000 scenarios, we simulated first an uncontrolled cases, wherein both the infectious and susceptible individuals are unmasked and ventilation rates are at their baseline level.  We then modified the simulations to include various combinations of controls --- a non-medical surgical face mask on the infectious person, a non-medical surgical mask on the susceptible person, and/or increases in the air change rate --- and evaluated the number of pathogens reaching the infection site relative to the uncontrolled case. We also quantified the impact of distancing on transmission by quantifying the risk of infection during a face-to-face interaction with 1~m or 2~m distancing relative to a close (0.5~m) conversation.

To simulate a susceptible person wearing a mask while an infectious person is not, we assumed that some of the particles, depending on their size, were captured in the mask; these captured particles are unable to deposit into the nasal cavity of the susceptible person and cause infection. To estimate the size-dependent collection efficiency upon inhalation, we used the mean and standard deviation from \cite{pan2021}, which are given in Table~\ref{tab:maskFE}. Using these distributions, we represented a separate mask collection efficiency curve for each scenario, where the mask efficiency curve was varied among scenarios to reflect the standard deviation of the measurements. To ensure that the collection efficiency curves in each scenario are monotonic with respect to particle size, we assumed the relative performance (in terms of standard deviations from the mean) was the same across particle sizes within each scenario. Similarly, we represented a face mask on the infectious person by modifying the number of virus-laden particles that are expelled, according to the scenario- and size-dependent mask collection efficiency upon exhalation, where the mask efficiency upon exhalation was also estimated from \cite{pan2021}. Following measurements by \cite{maher2020}, we assumed that wearing a face mask reduces the expiration velocity by 53\% ($\pm$12\%) \citep{maher2020}; we adjusted the effective diameter of the jet orifice to maintain the same flow rate as in the unmasked case. We focused on non-medical surgical masks only.%

To quantify changes in the risk of infection with ventilation, we simulated each scenario with its baseline (low) ventilation rate and at a ventilation rate elevated by four air changes per hour (ACH). While the absolute ventilation rates for the baseline and enhanced ventilation cases varied among scenarios, the difference between the ventilation rate in each case was held constant. The ventilation rate varied from 0.3 to 2.7~ACH in the baseline cases and from 4.3 to 6.7~ACH in the enhanced ventilation cases. The 4-ACH increase was chosen to approximately represent the impact of widely opening several windows \citep{howard2002}, though the true impact of natural ventilation on ACH is highly variable depending on the particular building and the local meteorology. Here we assume that indoor air is exchanged with outdoor air, though findings would be similar for ventilation systems that employ high-efficiency filters.

\begin{figure}[h!]
\centering
    \includegraphics{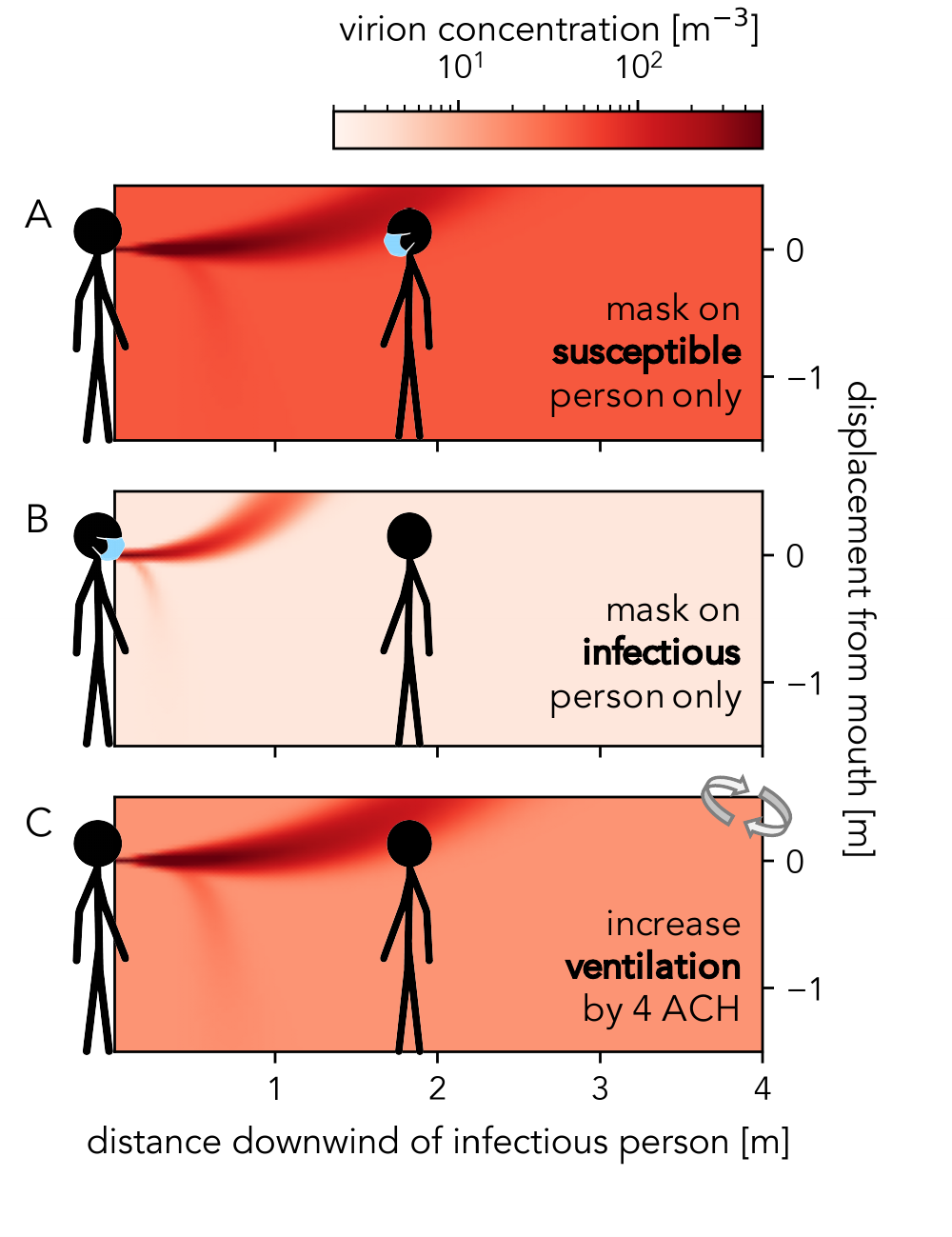}
 \vspace{-0.5cm}  
 \caption{Pathogen concentrations after a 3-hour, face-to-face conversation between an infectious person (left) and a susceptible person (right) in an example scenario. The susceptible person's exposure may be reduced if (A) they wear a mask, (B) the infectious person wears a mask, or (C) the ventilation rate is increased.
 \label{fig:C_xz}}
 \end{figure}

\section{Results}\label{sec:results}

\subsection{Face masks and ventilation}
 We first present the efficacy of face masks and ventilation in reducing initial infection of a susceptible person as a function of distance from an infectious person, then demonstrate the impact of combining these controls with social distancing in Section~\ref{sec:layers}. Before presenting the efficacy of controls across ensemble of simulations, we first illustrate our modeling approach using example scenario, which was the same as the baseline scenario in \cite{fierce2021IndoorAir}; parameters are specified in Table~1 of \cite{fierce2021IndoorAir}.

\begin{figure}
 \centering
 \vspace{0.2cm}
    \includegraphics{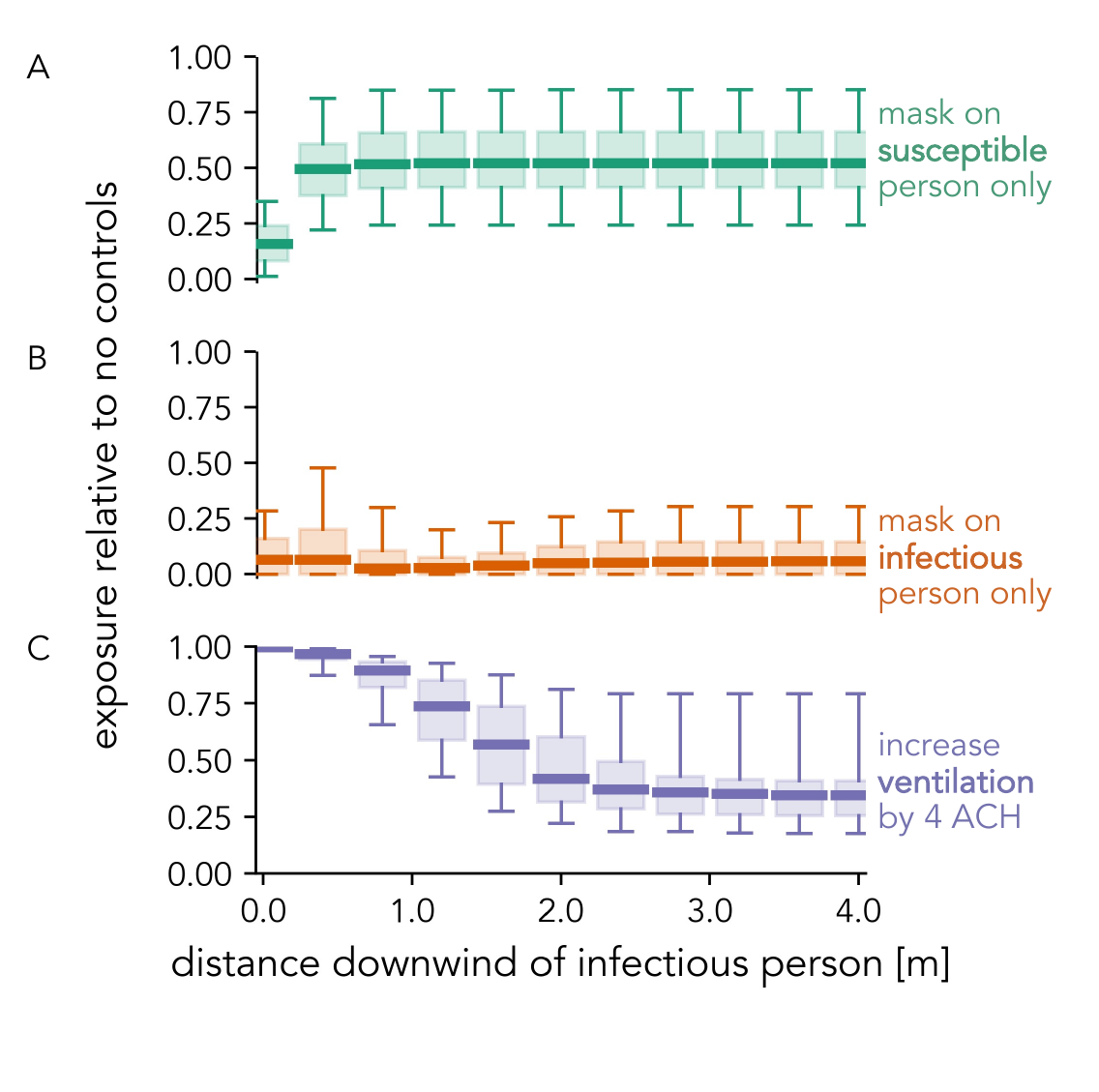}
 \vspace{-0.5cm}  
 \caption{Mean (solid line), quartiles (boxes), and 5\%--95\% confidence intervals (whiskers) of the exposure to airborne pathogens relative to no controls at varying distance from an infectious person. Shown for the same general scenario, but considering different individual controls: (A) the susceptible person wears a face mask and the infectious person does not, (B) the infectious person wears a face mask and the susceptible person does not, and (C) neither person wears a mask but ventilation is increased by 4 air changes per hour.}
 \label{fig:controls}
 \end{figure}

 If a susceptible person is in a room with an infectious person, they may be exposed to high concentrations of pathogens depending on their respective locations and the conditions within the room (see Fig.~\ref{fig:C_xz}a). When the susceptible person wears a mask, their exposure to pathogens is reduced because the mask captures some of the infectious particles. The median risk of infection at the same distancing is reduced by 50\% if the susceptible person is wearing a mask (Fig.~\ref{fig:controls}a). The relative exposure at any distance is highly variable, with 90\% confidence intervals ranging from a relative risk of nearly 0.25 (75\% reduction in risk) to 0.75 (25\% reduction in risk) at distances beyond 0.2~m. 
In comparison, we find that mask use by the infectious person generally has greater benefits than mask use by the susceptible person (comparison between Fig.~\ref{fig:controls}a~and~\ref{fig:controls}b), resulting from collection efficiencies upon expiration being larger than upon inhalation \citep{pan2021}. Mask use by the infectious person leads to a reduction in pathogens concentrations both in their expiratory jet and throughout the room (comparison between Fig.~\ref{fig:C_xz}a and~\ref{fig:C_xz}b). Wearing a face mask also reduces the velocity of expelled particles \citep{maher2020}, which reduces the horizontal extent of their expiratory jet \citep{fierce2021IndoorAir}. When the infectious person wears a face mask, the risk of initial infection in the susceptible person is reduced by approximately 90\%, and the efficacy showed lower variability than the efficacy of a face mask on the susceptible person. 

Whereas face masks reduce the relative risk both near and far from an infectious person, the efficacy of ventilation in reducing exposure to pathogens depends strongly on the distance between the infectious and the susceptible person, as shown in Figures~\ref{fig:controls}c. While many of the pathogens are carried in small particles that remain suspended and affect concentrations throughout the room \citep{li2021, miller2021, azimi2021}, concentrations are greatest in the region directly downwind of the infectious individual. If the infectious person is unmasked, but the ventilation rate is increased by 4~ACH (Fig.~\ref{fig:C_xz}c), far-field pathogen concentrations are reduced while concentrations in the expiratory jet of the infectious individual remain virtually unchanged. Consequently, increasing the ventilation rate does little to reduce the risk of near-field transmission. On the other hand, with distancing of 2~m or more, increasing the ventilation rate by 4~ACH reduces the median risk of transmission by more than 70\%.



\begin{figure*}
\centering
 \centering
 \includegraphics[scale=0.5]{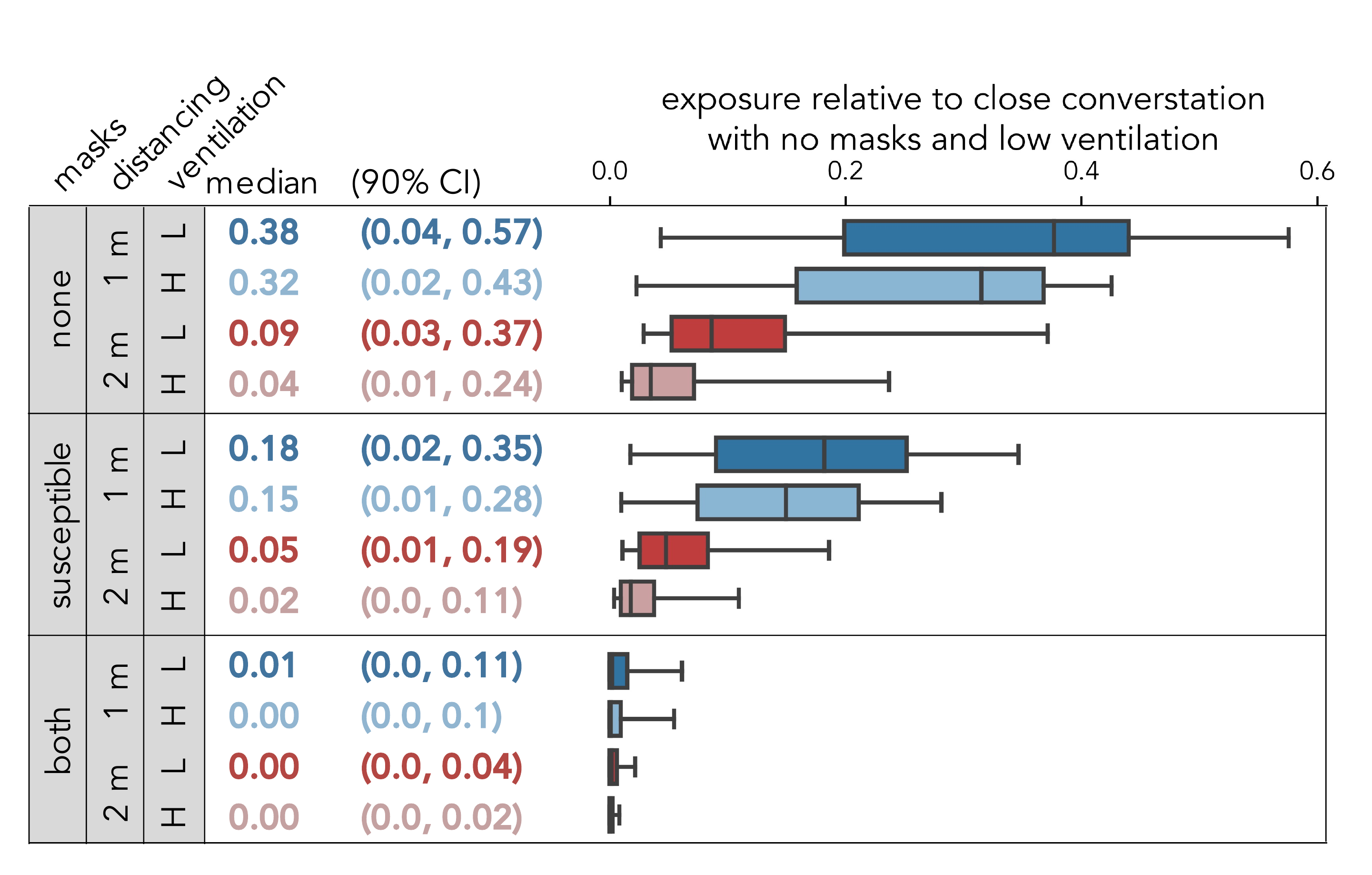} 
 \caption{Median (solid line), quartiles (boxes), and 90\% confidence interval (whiskers) of the fraction of airborne pathogens reaching the infection site when different combinations are employed in comparison with a close (0.5~m) conversation at the baseline (low) ventilation rates. The median and 90\% confidence intervals of the relative exposure are also shown in the adjoining table.}
 \label{fig:layers}
 \end{figure*}

\subsection{Layering controls}\label{sec:layers}
The greatest reductions in the risk of initial infection in a susceptible person are achieved by layering multiple controls. For each combination of controls, we quantified exposure to pathogens relative to a close (0.5~m), unmasked conversation with an infectious person at the baseline (low) ventilation rates (see Fig.~\ref{fig:layers}).

If both people are unmasked, increasing the distance between them from 0.5~m to 1~m reduces the median exposure by more than 60\% (dark blue in top panel of Fig.~\ref{fig:layers}), whereas an increase in distancing from 0.5~m to 2~m reduces the median risk of exposure by more than 90\% (dark red in top panel of Fig~\ref{fig:layers}). Additionally, with greater distancing between individuals, the impact of ventilation tends to be greater. Whereas increasing the ventilation rate by 4 ACH leads to only small changes in exposure when the individuals are distanced by only 1~m (comparison between dark and light blue in Fig.~\ref{fig:layers}), reduces the susceptible person's median exposure by more than 50\% when the individuals are distanced. The combination of 2-m distancing and an increase in the ventilation rate by 4~ACH leads to a reduction in the median risk of exposure by 95\% if both individuals are unmasked. 

If one or both individuals wear face masks in combination with distancing, the susceptible person's exposure to pathogens is further reduced. The median number of pathogens reaching the infection site during a conversation between masked individuals at least 1~m apart is reduced by 99\% relative to a close (0.5~m) conversation without masks. Furthermore, when both are masked, the difference in infection risk between 1~m and 2~m of distancing is smaller than when both are unmasked.

\section{Discussion and Conclusions}\label{sec:conclusions}
In this study, we showed that layered controls are highly effective in reducing transmission of airborne pathogens like SARS-CoV-2. The combination of social distancing, masking, and increasing ventilation led to a reduction in the median risk of infection by more than 99\%. We also show that the efficacy of an individual control depends on the other controls that are in use. For example, ventilation reduces exposure to pathogens only when used in combination with social distancing; a susceptible person's exposure to airborne pathogens during a close conversation is governed by pathogen concentrations in the expiratory jet, which is not strongly impacted by the ventilation rate, but ventilation does impact long-range exposure to pathogens. On the other hand, increasing distancing from 1~m to 2~m has a smaller impact when all parties are masked than when all are unmasked, which may explain the similarity in infection rates in schools with 3-foot and 6-foot distancing between masked students \citep{van2021}. Without universal masking, increasing distancing from 1~m to 2~m is critical for minimizing the risk of infection.

Although the pathogen dose to a susceptible person and the resulting risk of initial infection depends strongly on parameters that are not well constrained, vary among pathogens, and are impacted by vaccinations, through a global sensitivity analysis we found that variability in the efficacy of controls is driven by variability the mask characteristics and room conditions among scenarios. To identify the parameters that most affect variability in efficacy of an individual control, we used the Sensitivity Analysis Library in Python, which is available at \url{https://salib.readthedocs.io/en/latest/.} Unsurprisingly, the impact of increasing ventilation tends to be greatest in small, poorly ventilated spaces; the global sensitivity analysis revealed that variability in the efficacy of ventilation is controlled by the baseline ventilation rate and the volume of the room. On the other hand, the efficacy of face masks was driven predominantly by the overall mask performance, which varies with filtration efficiency of the mask material and leaks due to improper mask fit. This finding suggests that mitigation strategies should emphasize the use of high-quality, well-fitting masks. Measurements show that surgical masks tend to have higher collection efficiencies than microfiber or cotton masks \citep{pan2021, asadi2020}, while using a surgical mask with a cotton mask is even more effective \citep{drewnick2020}. Masks can be further adjusted to improve fit \citep{blachere2021face}. Reductions in risk will likely be greater than reported in this study if either party is wearing an N95 respirator \citep{asadi2020} or layering a well-fitting cotton mask on top of a non-medical surgical mask \citep{brooks2021}.

While the present modeling study is intended to provide mechanistic insights into the impact of nonpharmaceutical interventions on airborne transmission of pathogens, the modeling framework is intentionally idealized. The current implementation of the model does not account for complex air circulation patterns within indoor spaces, which may lead to spatial gradients in pathogen concentrations. Further, we do not represent the potential impact of enhanced air circulation on the expiratory jet of an infectious person. We also focus on the impact of controls on transmission during a face-to-face conversation between an infectious person and an susceptible person; we do not consider the potential impact of leakage from the side of masks on people standing beside an infectious person.

In summary, we showed that the efficacy of any individuals controls was highly variable among scenarios. However, even with these large uncertainties we find that employing a combination of distancing, universal masking, and enhanced ventilation reduced the risk of initial infection by more than 98\% in 95\% of cases, suggesting that layered nonpharmaceutical interventions offer high levels of protection, which can be applied in combination with vaccines. Wide adoption of layered controls could dramatically reduce transmission of existing airborne pathogens, such as SARS-CoV-2, and will be critical for controlling outbreaks of novel pathogens in the future.

\section*{Acknowledgements}
This research was supported by the DOE Office of Science through the National Virtual Biotechnology Laboratory, a consortium of DOE national laboratories focused on response to COVID-19, with funding provided by the Coronavirus CARES Act. This project was supported in part by the U.S. Department of Energy through the Office of Science, Office of Workforce Development for Teachers and Scientists (WDTS) under the Science Undergraduate Laboratory Internships Program (SULI). The quadrature-based model was originally developed with support from the DOE Atmospheric System Research program.

\section{Data Availability}
The QuaRAD source code, input files, and processing script are available for download at: \url{https://github.com/lfierce2/QuaRAD/}. Simulation ensembles were created using latin hypercube sampling with pyDOE: \url{https://pythonhosted.org/pyDOE/}. The sensitivity analysis was performed using the Sensitivity Analysis Library in Python, which is available at: \url{https://salib.readthedocs.io/en/latest/}.

\section*{Conflict of Interests}
The authors declare no conflicts of interest.


\bibliography{references}

\end{document}